\documentstyle[epsfig,11pt]{article}
\def\be{\begin{equation}}
\def\ee{\end{equation}}

\begin{document}
\begin{titlepage}

\mbox{}
\vspace{4cm}

\begin{center}
{\bf\LARGE On $\pi_0 \rightarrow \gamma\gamma$ and the 
axial anomaly at $T\not= 0$ $^{\dagger}$}
\vspace{2cm}

\centerline{\bf 
                Robert D. Pisarski and 
Michel H.G. Tytgat $^{\ast}$}
\vspace{0.6cm}

\centerline{Physics Department, Brookhaven National Laboratory,}
\centerline{Upton, New York 11973-5000, USA}

\vspace{1cm}
\centerline{May 1997}

\vspace{3cm}

{\bf Abstract}\\
\parbox[t]{\textwidth}
{In  vacuum,
in the chiral limit
the coupling of a pion to two on-shell
photons is directly related to the coefficient of the axial
anomaly in QED.  This relationship is lost at any nonzero temperature.
Explicit calculations show  that
the coupling decreases with temperature and  vanishes at $T_c$, the
temperature of chiral symmetry restoration.}
\end{center}
\vspace{\fill}

{\noindent\makebox[10cm]{\hrulefill}\\
\footnotesize
\makebox[1cm][r]{$^{\dagger}$}Talk given at the XXXIst Rencontres de Moriond: QCD and 
Strong Interactions.

\noindent
\footnotesize
\makebox[1cm][r]{$^{\ast}$}mtytgat@wind.phy.bnl.gov
}
\end{titlepage}
\newpage

Consider QCD with two light flavors ({\it up} and {\it down}).
The Lagrangian has an approximate $SU(2)_l \times SU(2)_r$
chiral symmetry, spontaneously
broken to $SU(2)$ in vacuum. The resulting Goldstone bosons (the pions) 
have a direct coupling to the axial current, 
\be
\label{eq1}
\langle 0 \vert J_{\mu 5}^a \vert \pi^b(P) \rangle = i P_\mu f_\pi
\delta^{a b} 
\ee
with $a, b = 1..3$ and $f_\pi \sim 93 MeV$ is the pion decay constant.
In 
the chiral limit the axial current is
conserved and, consequently, the pions are massless. 

Currents which are conserved classicaly may not remain so 
quantum mechanically.
In particular, at one-loop order the third component of the
axial current develops an anomalous divergence \cite{adler},
\be
\label{eq2}
\partial_\mu J_{\mu 5}^3 = - {\alpha_{em} \over 8 \pi}
\varepsilon_{\mu\nu\alpha\beta} F_{\mu\nu} F_{\alpha \beta}
\ee
where $F_{\mu\nu}$ is the QED field strength.
A  striking manifestation of the QED axial anomaly is to allow the
$\pi^0$ to decay into two photons. Let us define the amplitude for
$\pi^0 \rightarrow \gamma\gamma$ as
\be
{\cal A} = g_{\pi\gamma\gamma}\,\varepsilon_{\alpha\beta\mu\nu}
\,\epsilon^1_\alpha \epsilon_\beta^2\,k^1_\mu k_\nu^2
\ee
where $k^1$, $k^2$, and $\epsilon^1$, $\epsilon^2$ are the
momenta and polarization vectors of the photons. 
In vacuum, the spontaneous breaking of chiral symmetry, 
~(\ref{eq1}), and the anomaly equation, ~(\ref{eq2}),
suffice to deduce the beautiful relation,
\be
\label{eq4}
f_\pi\, g_{\pi \gamma\gamma} = {1\over \pi}\, \alpha_{em}.
\ee
This formula is 
valid in the {\em chiral limit} and for {\em on-shell photons}.
The {\em r.h.s.} of~(\ref{eq4}) comes from the anomaly
of~(\ref{eq2}): without this term, $g_{\pi \gamma\gamma}$ would
vanish, which is the content of the Sutherland-Veltman theorem~\cite{velt}. 
Another remarkable theorem~\cite{adler} states that the coefficient of
the anomalous term  is not affected by higher-order
corrections, so that in the vacuum, ~(\ref{eq4}) is exact.

What about finite temperature ? The non-renormalization of the axial
anomaly holds at finite T or density~\cite{mueller,ano2} but
is~(\ref{eq4}) still valid ? 
Let us first assume so. We know~\cite{fpit} that, to leading order at
low temperature and
in the chiral limit, the pion decay constant changes as
\be
\label{eq5}
f_\pi(T) = f_\pi \left ( 1 - {T^2\over 12 f_\pi^2}\right ).
\ee
(for $N_f = 2$ light flavors).
As the {\em r.h.s.} of~(\ref{eq4}) is $T$
independent~\footnote{We neglect $\delta \alpha_{em}(T) \sim \log(T)$,
corrections. These are subleading with respect to
the  $T^2$ term of $f_\pi(T)$ in~(\ref{eq5}).}, one concludes
that $g_{\pi\gamma\gamma}(T)$ {\em increases} at low $T$,
$g_{\pi\gamma\gamma}(T) = g_{\pi\gamma\gamma}(1 + T^2/ 12
f_\pi^2)$. At higher temperatures,  $f_\pi(T)$ goes 
to zero at  $T_c$, where  chiral symmetry becomes manifest, which
would imply that
$g_{\pi \gamma\gamma}$ blows up at $T_c$. 
This is obviously absurd and~(\ref{eq4}) cannot hold as such at finite
temperature. Actually, a closer look at the conditions under
which~(\ref{eq4}) is derived, reveals that Lorentz invariance is
crucial. In the presence of a thermal bath, Lorentz invariance is lost
and~(\ref{eq4}) is not true at any non-zero temperature. 
Apparently, this has not been fully appreciated before
(see~\cite{mueller} for related issues). 
A more detailled
discussion of our claim based on Ward identities can be found in~\cite{pistyt}.

How is  $g_{\pi\gamma\gamma}$ changing with temperature ?
Because the direct connection with the coefficient of the anomaly is
lost, this is a dynamical problem that can only be adressed by
an explicit calculation. In this respect, finite T is
similar to the problem of computing the amplitude for $\pi \rightarrow
\gamma\gamma$ for {\em off-shell} photons (see {\em
e.g}~\cite{leader}):
 the Ward identities provide
only limited information . 

Consider first a linear sigma model with constituent
quarks (see~\cite{rob,itz} for details). This is a renormalizable
theory and the pion has a
pseudoscalar coupling to the quarks~\footnote{We consider a simple
$U(1) \times U(1)$ model. The extension to the non-abelian case is trivial.},
\be
{\cal L}_{int} = g \phi \bar q_R q_L + h.c. = g\, 
\sigma \bar q q + i g\, \pi \bar q \gamma_5 q
\ee
The scalar potential is such  that the chiral symmetry is
spontaneously broken in vacuum, 
$\sigma \rightarrow \sigma_0 + \sigma$. Then, to
leading order, 
$f_\pi = \sigma_0$ and $m_q = g \sigma_0$ and the $\pi$ is a Goldstone
boson. The $\pi$ couples to two photons through the  triangle
diagram (fig. 1 below). Because the pion-quark coupling is pseudoscalar,
$g \gamma_5$, 
and the photons vertices are vector-like, a mass insertion is
necessary to restore chirality; thus Dirac trace brings in
one power of the constituent quark mass.  The remaining
loop integral is ultraviolet finite, 
\begin{equation}
\label{eq7}
g_{\pi\gamma\gamma} \propto \alpha_{em} \, g \, m_q \int\! d^4\!p\, {1\over
(p^2 + m_q^2)^3
}
\ee
The constituent quark mass provides and infrared cut-off, so
$I \propto 1/m_q^2$, and
\be
g_{\pi\gamma\gamma} = {\alpha_{em} \over \pi} g {m_q \over m_q^2} =
{\alpha_{em} \over \pi} \, {1\over f_\pi} \; .
\ee
Thus, as expected, the constituent quark model respects 
~(\ref{eq4}).

\begin{figure}[hbt]
\centerline{\epsfig{figure=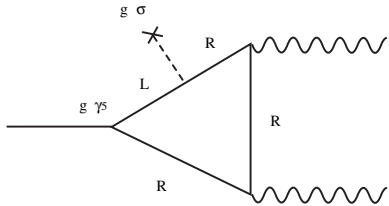,height=28mm}}
\caption{$\pi_0$ couples to
$\gamma\gamma$ through the triangular diagram. $L$ and $R$ refer to  quark chiralities.}
\end{figure}

It is easy to extend this calculation to finite $T$. As there
is no dependence in the external momenta,
we  compute
the integral $I$ in  the imaginary time formalism. Then 
\be
I \rightarrow I_T = T \sum_{n=-\infty}^{+ \infty} 
\int d^3\!p {1\over (p^2 + m_q^2)^3} 
\ee
with $q_0 = \pi(2 n+1 ) T$. As $T$ gets close to $T_c$, $m_q \ll T$,
and $T$ provides the infrared cut-off, 
$I_T \propto 1 / T^2$.
Consequently, near and below $T_c$, 
\be
\label{eq10}
g_{\pi\gamma\gamma}(T) \propto \alpha_{em} \, g^2 {f_\pi(T) \over T^2}
\rightarrow 0
\ee
This calculation suggests that the coupling goes to zero at $T_c$ as a
consequence of chiral symmetry restoration. For further arguments,
see~\cite{rob}.

Now, let us  consider the low temperature limit. 
At low $T$, the dynamics is dominated by the massless pions 
and we use a gauged non-linear sigma model
with a Wess-Zumino term~\cite{witten},
\begin{eqnarray}
\label{eq11}
{\cal L} &=& {1\over 2} \,(D_\mu \pi^a)^2 + {1\over 6
f_\pi^2} (\pi^a \partial_\mu \pi^a)^2 + \ldots\nonumber\\
&& \mbox{} + \left( {e^2 N_c\over 96 \pi^2} \right) \, {1 \over f_\pi}\,
\pi^0 \varepsilon_{\alpha\beta\mu\nu} F_{\alpha \beta} {F}_{\mu \nu} + \ldots
\end{eqnarray}
where the dots stands for operators with more pions  or
higher derivatives. At tree level, the effective
Lagrangian~({\ref{eq11}) is normalized so that~(\ref{eq4}) holds~\cite{witten}.
Pion one-loop corrections (fig. 2) to the anomalous vertex  are easily
computed. Consider first vacuum corrections~\cite{don}. In the
chiral limit and for on-shell photons, these simply amount to 
 replace $f_\pi$ in~(\ref{eq11}) by its renormalized, physical
value. This is how the Adler-Bardeen theorem, which states
that the anomaly is not renormalized beyond one loop order,
works at the level of the effective action.
In the chiral limit $m_\pi \rightarrow
0$ and for on-shell photons, the ${\cal O}(P^4)$ 
anomalous operator of~(\ref{eq11}) is
the unique operator which contributes to 
the amplitude for $\pi_0 \rightarrow
\gamma\gamma$; all other operators are  ${\cal O}(P^6)$ or
higher, and vanish on the photon mass shell. 
Hence, all the  divergences have to  be
absorbed by (\ref{eq11}).

\begin{figure}[hbt]
\centerline{\epsfig{figure=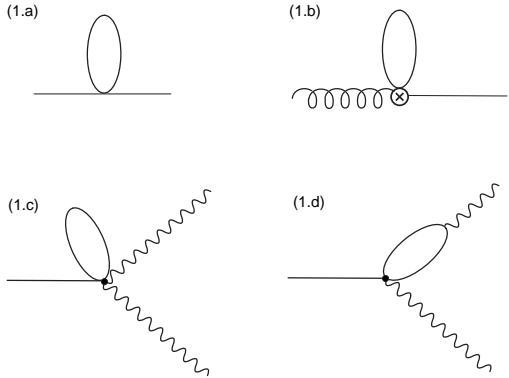,height=50mm}}
\caption{pion one-loop corrections to $\pi_0\rightarrow
\gamma\gamma$ in the gauged WZW model.} 
\end{figure}

At finite temperature, $T\ll f_\pi$, a careful 
calculation~\cite{pistyt} reveals \footnote{The computation of
$g_{\pi\gamma\gamma}$ in the  low T regime has been first addressed  by A. Gomez
{\it et al} \cite{estrada} who considered also the case $m_\pi \not= 0$.}
that
\be
\label{eq12}
g_{\pi\gamma\gamma}(T) = g_{\pi \gamma\gamma} \, \left ( 1 - {T^2
\over 12 f_\pi^2}\right)
\ee
Hence, $g_{\pi\gamma\gamma}$ decreases  like
$f_\pi(T)$ at low temperature~(\ref{eq5}), consistent 
with~(\ref{eq10}) close to $T_c$. 
 That the temperature dependence of $g_{\pi\gamma\gamma}$ can be  
non-trivial is due  to the
fact that at $T\not= 0$, 
unlike in the vacuum, it is possible
to add new ${\cal O}(P^4\, T^2/f_\pi^2)$ terms to the
 Lagrangian~(\ref{eq11})~\cite{pistyt,pistyt2}. 
These terms are non-local, similar to the hard thermal loops
of hot QCD ~\cite{htl}.

The result~(\ref{eq10}) shows that the amplitude for $\pi\rightarrow
\gamma\gamma$
vanishes at the critical temperature. With non-zero
quark current quark masses, this  translates to a 
decrease  of the effective coupling near $T_c$, with  $f_\pi(T_c) \sim 1/3
f_\pi$. 
Also, in the linear sigma model,  if
$\pi\rightarrow \gamma\gamma$ is suppressed, $\pi \sigma \rightarrow
\gamma\gamma$ is not~\cite{rob}. This is because the  mass insertion,
$\propto \sigma_0$, is equivalent to the insertion  of a  $\sigma$ particle.

\bigskip 

This work is supported by a DOE grant at 
Brookhaven National Laboratory, DE-AC02-76CH00016.

\end{document}